\documentclass[aps,prl,showpacs,amsmath,amssymb,twocolumn,nofootinbib]{revtex4}
\usepackage{pslatex,graphicx,dcolumn,bm,natbib,amsmath,amsbsy}
\date{\today}
\usepackage{epsfig}
\usepackage{color}

\newcommand{\subfig}[1]{({\it#1\/})}
\newcommand{\pic}[2]{\begin{picture}(0,#1)(0,0)#2\end{picture}}
\newcommand{\putpic}[2]{\put(#1){\begin{picture}(0,0)(0,0)#2\end{picture}}}

\begin{document}

\title{Experimental Demonstration of Nonuniform Frequency Distributions of
Granular Packings}

\author{Guo-Jie Gao$^{1}$, Jerzy Blawzdziewicz$^{1,2}$, Corey S. O'Hern$^{2}$ and Mark Shattuck$^{3}$\\
\normalsize{$^{1}$ Department of Mechanical Engineering, Yale University, New Haven, CT 06520-8286\\
$^{2}$Department of Physics, Yale University, New Haven
CT 06520-8120\\
$^3$Benjamin Levich Institute and Physics Department,
The City College of the City University of New York, New York, NY  10031}}

\begin{abstract}
We developed a novel experimental technique to generate mechanically
stable (MS) packings of {\it frictionless} granular disks.  We
performed a series of coordinated experiments and numerical
simulations to enumerate the MS packings
in small 2D systems composed of bidisperse disks.  We find that
frictionless MS packings occur as discrete, well-separated points in
configuration space and obtain excellent quantitative agreement
between MS packings generated in experiments and simulations.  In
addition, we observe that MS packing probabilities can vary by many
orders of magnitude and are robust with respect to the
packing-generation procedure.  These results suggest that the most
frequent MS packings may dominate the structural and mechanical
properties of granular systems.  We argue that these results for small
systems represent a crucial first-step in constructing a statistical
description for large granular systems from the `bottom-up'.
\end{abstract}
\maketitle

The power of equilibrium statistical mechanics is that it enables
the evaluation of macroscopic state variables (such as temperature and
pressure) of a macroscopic system in thermal equilibrium simply by
counting microstates.  A number of recent studies have applied similar
statistical methods to describe dense granular materials
\cite{bulbul,makse,coniglio}.  For example, Edwards-ensemble descriptions are
based on an assumption that all mechanically stable (or `jammed')
configurations of a granular system under a given set of macroscopic
constraints are equally likely.  The Edwards' theory also implies the
existence of a temperature-like variable---the compactivity $\chi$
\cite{edwards}.   

%%%%%%%%%%%%%%%%%
\begin{figure}

\pic{220}{

%TOP PANELS

\putpic{-120,105}{

\put(10,0){\includegraphics[width=0.15\textwidth]
                {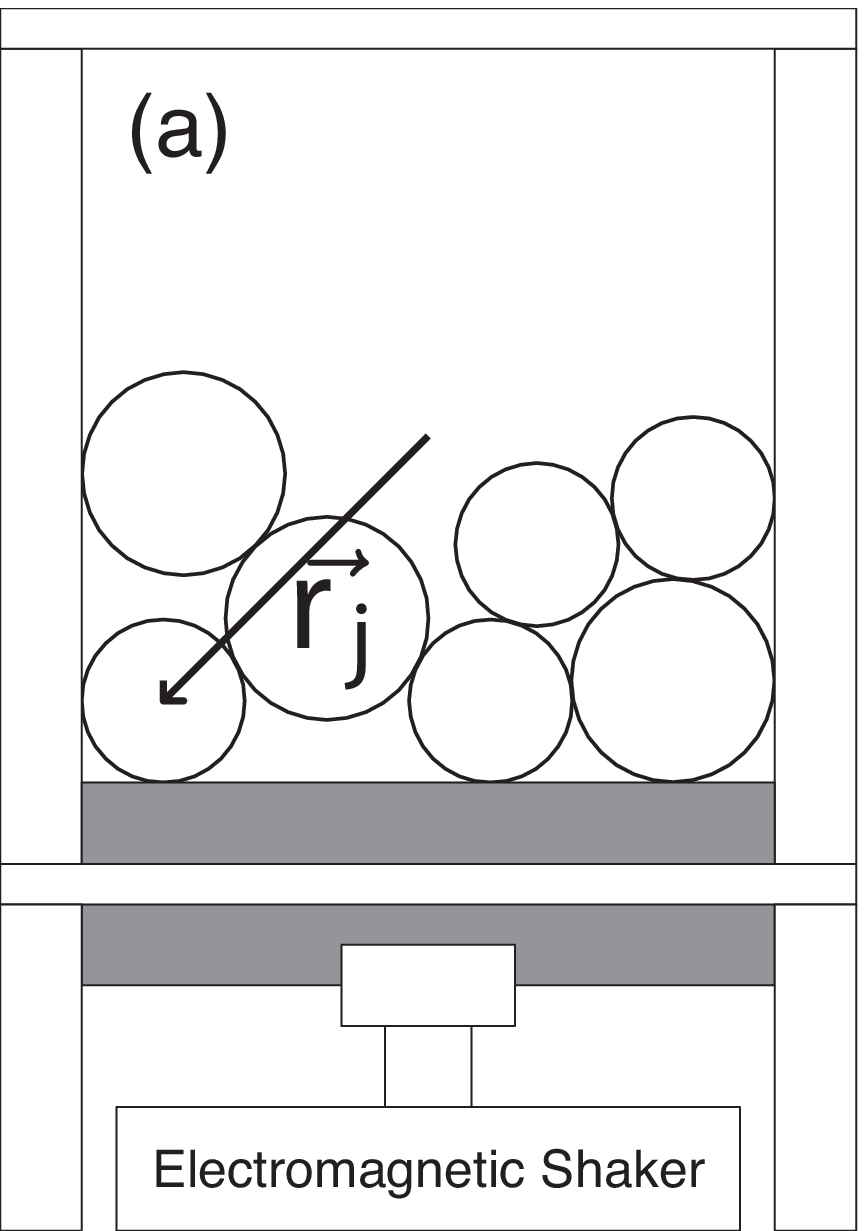}}

\putpic{100,10}{\includegraphics[width=0.28\textwidth]
                {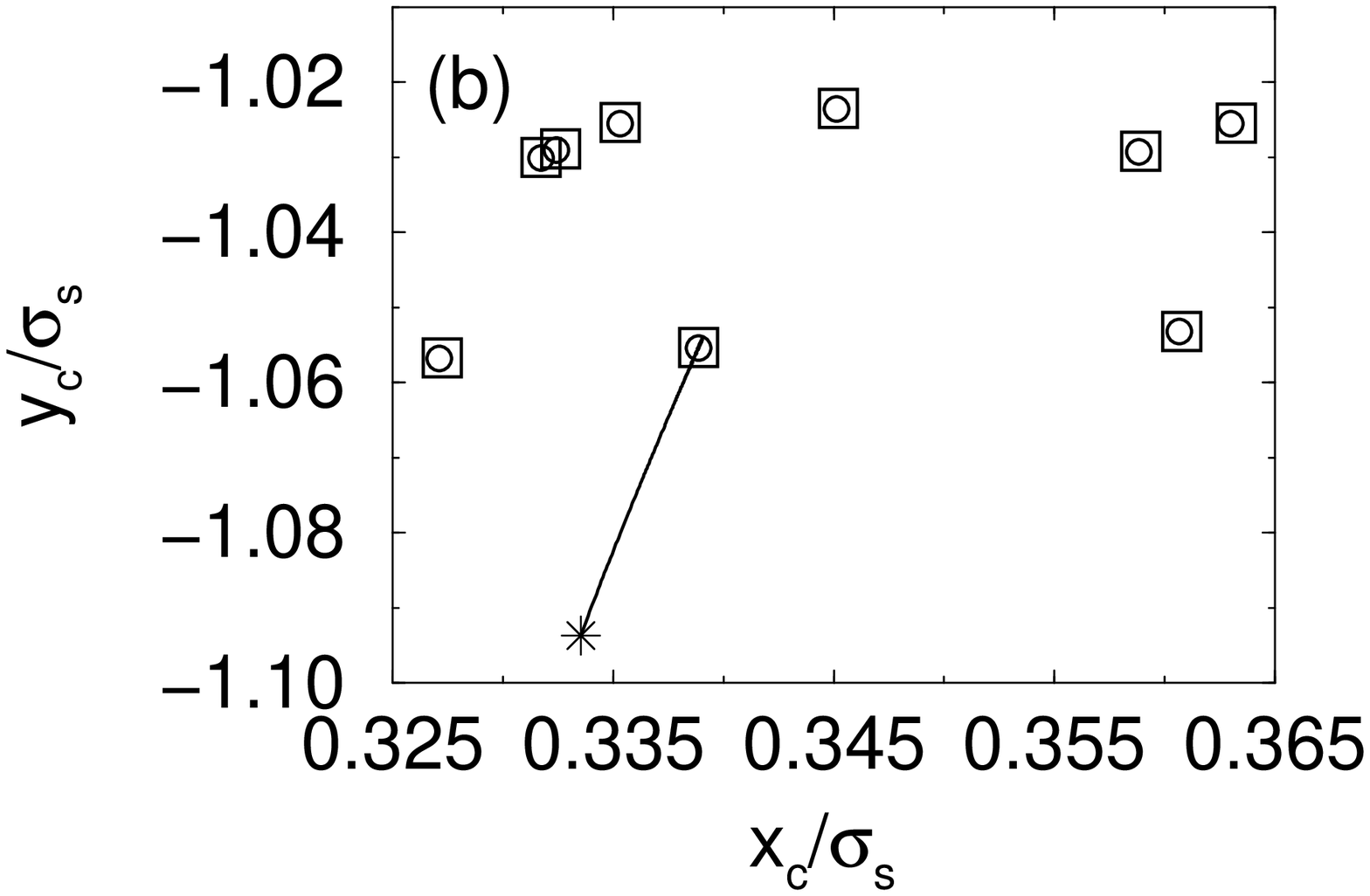}}

}

%BOTTOM PANEL

\putpic{-125,0}{

\put(0,10){\includegraphics[width=0.22\textwidth]
                {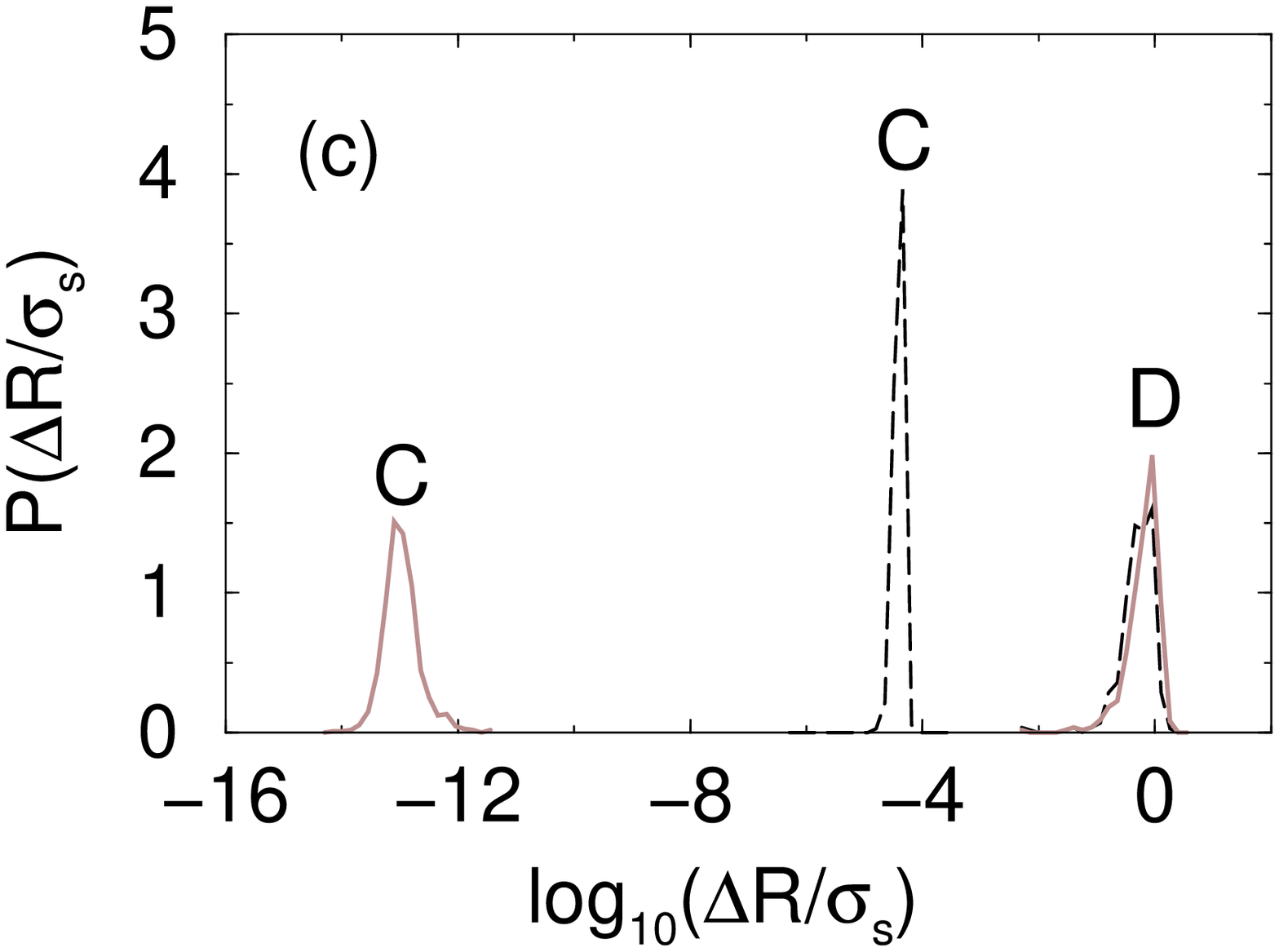}}

\put(127,10){\includegraphics[width=0.22\textwidth]
                {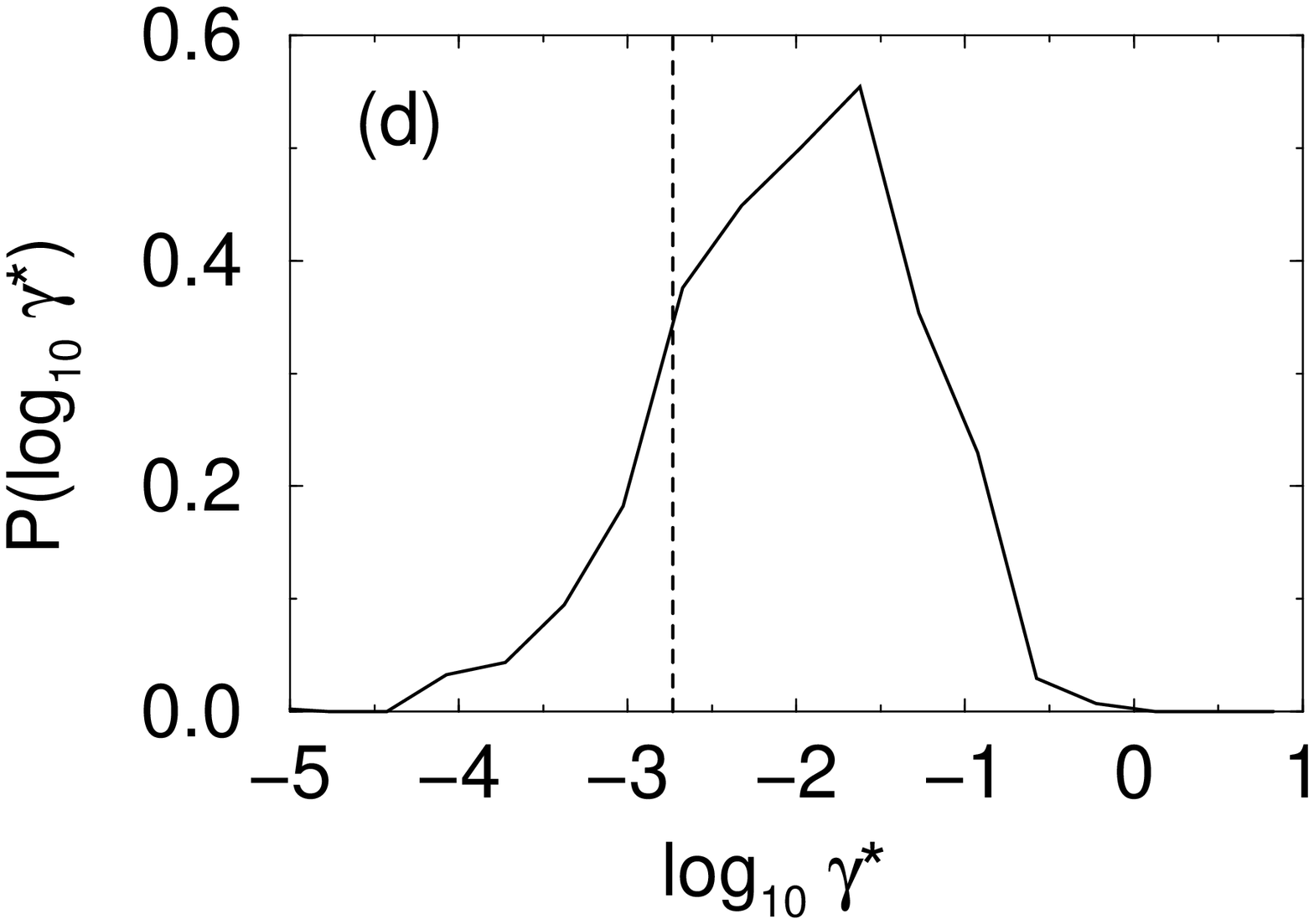}}

}
}
\vspace{-0.25in}
\caption{\label{experiment and state resolution} \subfig{a} Schematic
of experiment.  \subfig{b} Coordinates $(x_c,y_c)$ of the centroids of
several $N=7$ MS packings from experiments on plastic (squares) and
simulations for $\gamma_{\rm plastic}$ (circles). The solid line shows
the location of one of the centroids for
$10^{-7}\le\gamma\le\gamma^*$, where $\gamma^*$ is the value at which
the MS packing becomes unstable.  The star indicates the location of
the centroid at $\gamma^*$. \subfig{c} Probability distributions of
the separation $\Delta R$ in configuration space between distinct MS
packings (D) and between a given MS packing and the packing furthest
away with the same contact network (C) for experiments (dashed lines)
and simulations (solid lines). \subfig{d} Probability distribution for
$\gamma^*$ from simulations. The vertical line indicates
$\gamma=\gamma_{\rm plastic}$.}
\vspace{-0.25in}
\end{figure}
%%%%%%%%%%%%%%%%%  

Despite the fact that granular media are dissipative and require
external driving forces (not thermal fluctuations) to explore
configuration space, there has been surprising success in describing
these materials using statistical methods based on the Edwards'
assumption \cite{compaction}.  For example, simulations of
slowly sheared granular systems have shown that $\chi$ can be used to
quantify effective `thermal' equilibrium, since different tracer
particles achieve the same $\chi$ \cite{makse2} and several
equilibrium measures of temperature all agree \cite{ohern}.
However, statistical mechanics approaches for dense granular materials
have been applied without directly testing the underlying fundamental
assumptions.  In particular, the assumption of equal micrsostate
probability has not been tested explicitly, and the relevant
microstates have not been clearly defined.  We advocate a novel
`bottom-up' approach to constructing statistical mechanics descriptions
of dense granular materials---one where we enumerate the microstates
and accurately measure the probabilities with which they occur.

To do this, we performed a coordinated set of experimental and
computational studies of mechanically stable (MS) packings in small 2D
granular systems undergoing vertical vibrations.  To enable
enumeration of MS packings (i.e., microstates), we focused on
small systems with no frictional forces.  We show that in
the absence of frictional forces, the set of MS packings is {\it
discrete\/}; thus packing probabilities can be directly evaluated by
counting the frequency with which they occur in a long sequence of
independent trials.  In contrast, frictional MS packings
\cite{silbert} form continuous families, and therefore packing
probabilities cannot be uniquely determined without first defining an
appropriate probability measure.  To generate frictionless packings in
our experiments, we have developed a novel technique where frictional
forces are relaxed using small-amplitude, high-frequency vibrations.

In both experiments and simulations we find the following four key
results concerning the microstate distributions of granular packings
in the zero-friction limit: 1) There exist a finite number of discrete
MS packings that grows exponentially with system size.  2) The
frequency with which these packings occur is highly {\it nonuniform}.
3) The sets of packings found in experiments and simulations of model
granular media are very similar.  4) The packing frequency is
relatively insensitive to the packing preparation protocol.  We argue
that the above important new results need to be incorporated into
statistical descriptions of dense granular media.

%%%%%%%%%%%%%%%%%
\begin{figure}

\pic{355}{

\putpic{-120,40}{

\put(40,-18){\includegraphics[width=0.28\textwidth]
                {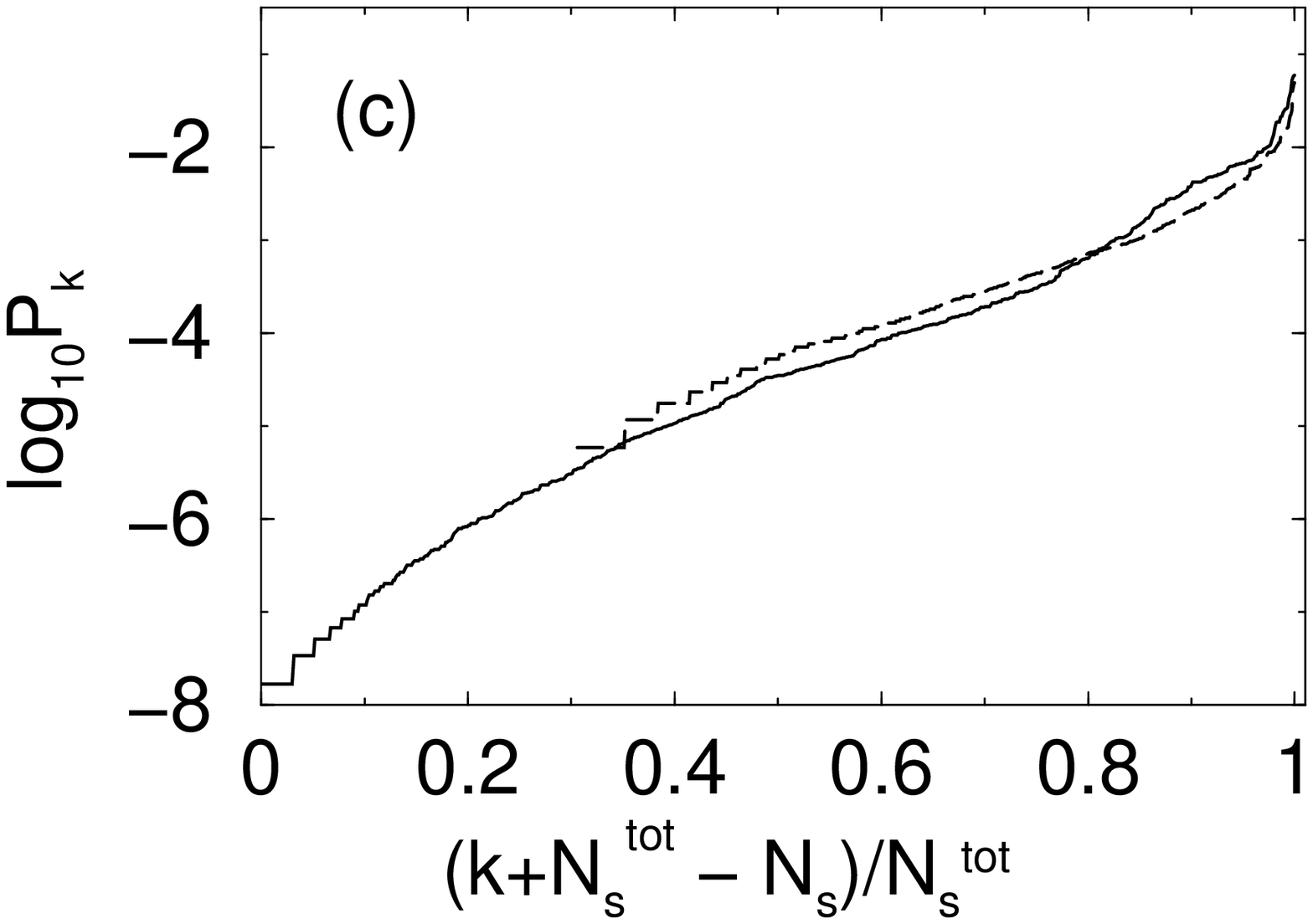}}

\put(40,90){\includegraphics[width=0.28\textwidth]
                {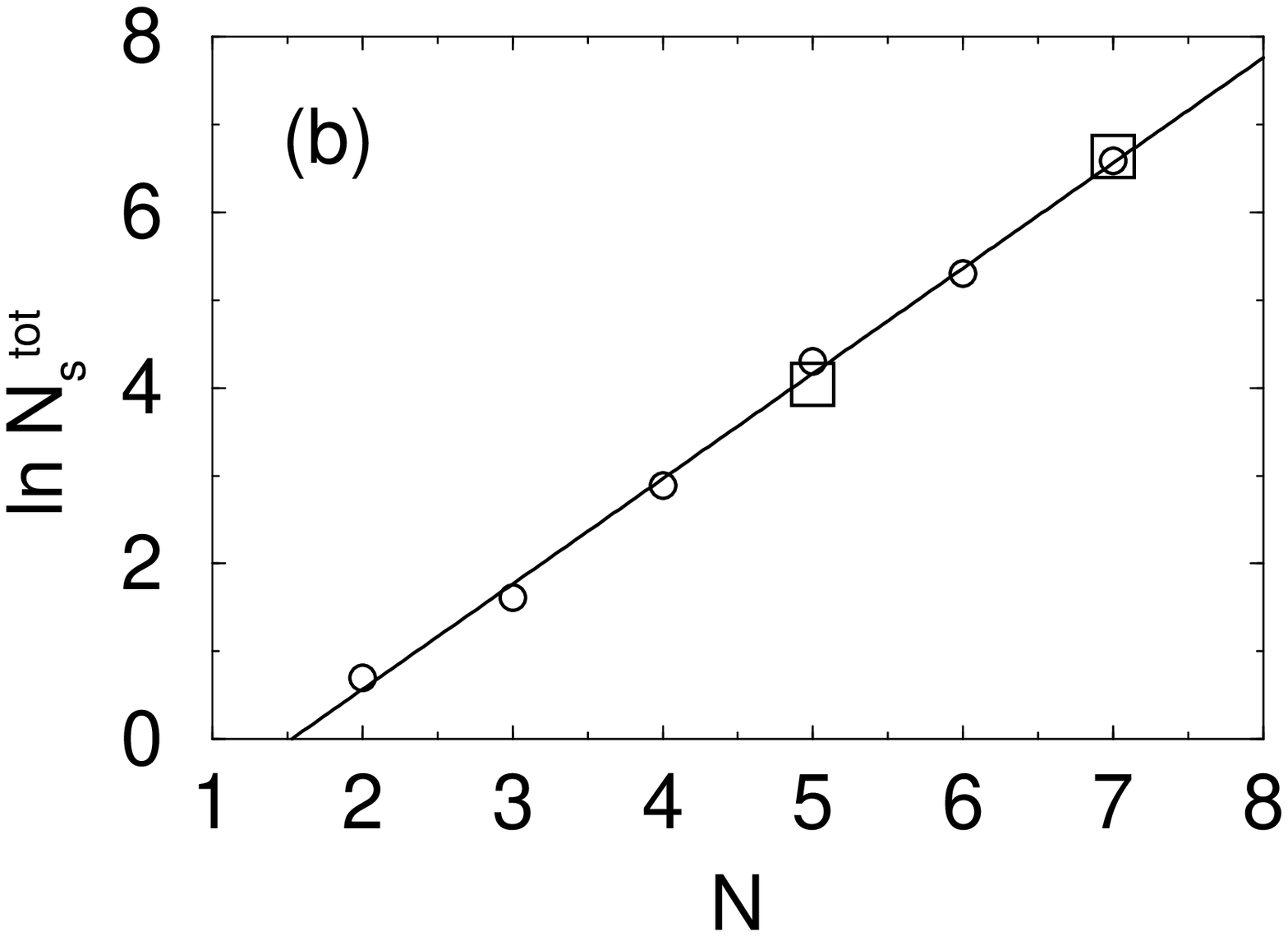}}

\put(40,200){\includegraphics[width=0.28\textwidth]
                {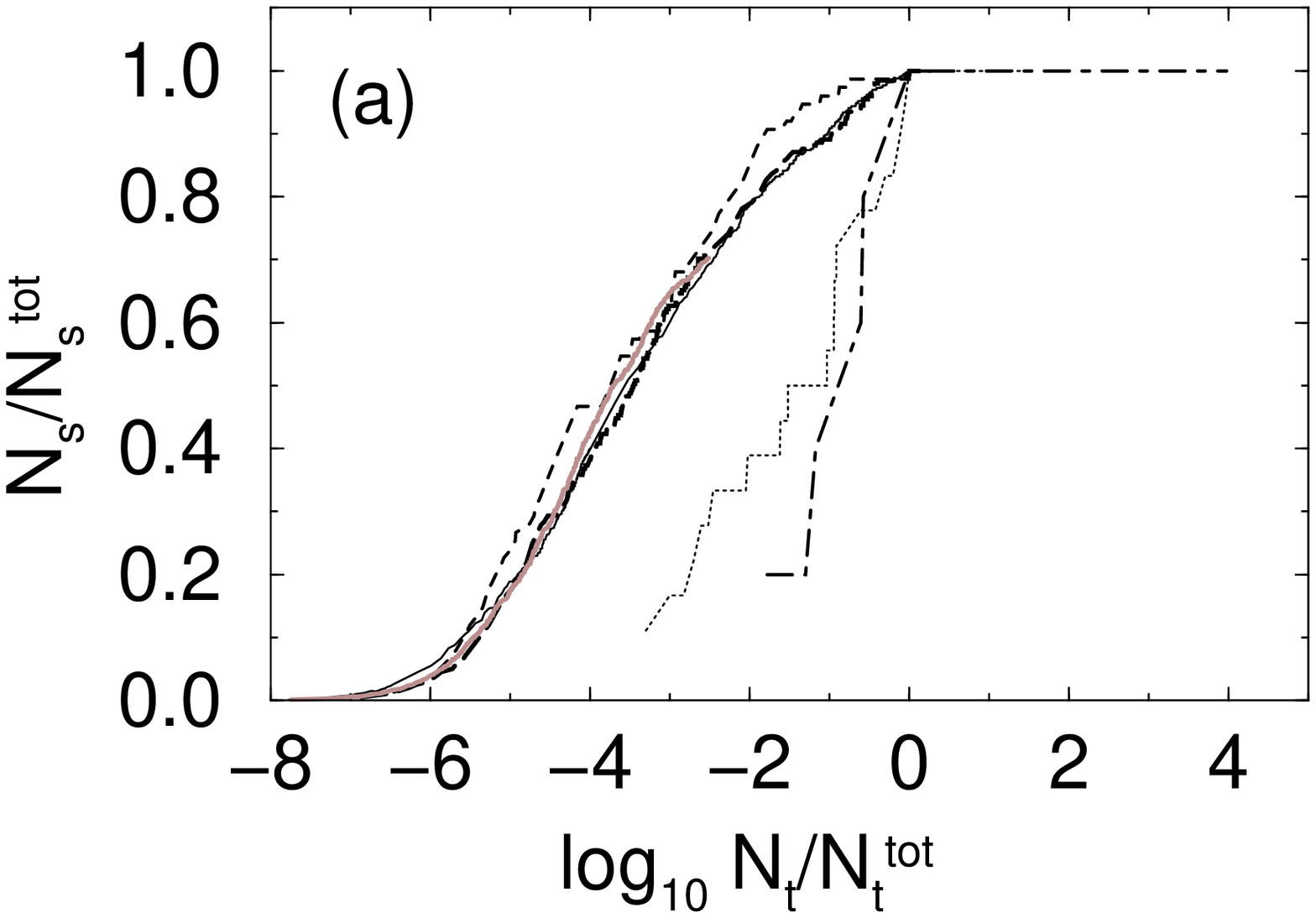}}
}
}
\vspace{-0.4in}
\caption{\label{packing probabilities} \subfig{a} Number of distinct
states $N_s$ found in $N_t$ trials for experiments on plastic (gray
solid line) and $N=3$ (dot dashed), $4$ (dotted), $5$ (dashed), $6$
(long dashed), and $7$ (solid) simulations at $\gamma=\gamma_{\rm
plastic}$.  The horizontal (vertical) axis is scaled by the total
number of MS packings $N_s^{\rm tot}$ (trials, $N_t^{\rm tot}$) at
saturation.  The experimental curve was obtained by fitting $N_s^{\rm
tot}$ and $N_t^{\rm tot}$ to simulations.  \subfig{b} $N_s^{\rm tot}$
vs. $N$ from simulations (circles) and experiments (squares). The
solid line has slope $1.2$. \subfig{c} Sorted probability $P_k$
of MS packings for $N=7$ vs. index $(k+N_s^{\rm
tot}-N_s)/N_s^{\rm tot}$ for simulations (solid) and experiments
(dashed). $N_s/N_s^{\rm tot} \approx 0.695$ for experiments was
determined from \subfig{a}.  }
\vspace{-0.22in}
\end{figure}
%%%%%%%%%%%%%%%%%  

\paragraph{Experiments}
A schematic of the apparatus used to generate MS frictionless disk
packings is shown in Fig.~\ref{experiment and state
resolution}\subfig{a}. A mixture of thin disks of thickness
$3.175\pm0.003$ mm and two different diameters $\sigma_s$ and
$\sigma_l$ were confined between two glass plates separated by
$3.20\pm 0.01$ mm and rested on a thin plunger connected to an
electromagnetic shaker through a slot in the bottom of the cell.  The
shaker enabled us to apply vertical vibrations at variable amplitude
and frequency to repeatedly generate static particle packings.

The particle mixtures consisted of $(N+1)/2$ small and $(N-1)/2$ large
particles, with $N=5$ and $7$.  For these systems we enumerated
the majority of frictionless MS packings by performing $N_t =
1.8\times 10^5$ ($N=7$) and $1.2\times 10^4$ ($N=5$) independent
trials using the protocol described below.  The ratio of disk
diameters was $d=\sigma_l/\sigma_s=1.2520 \pm 0.0003$, and the ratio of
the cell width $L$ to the small particle diameter was $\lambda =
L/\sigma_s = 4.25314 \pm 0.00001$ ($2.65 \pm 0.02$) for $N=7$ ($5$).
We used bidisperse systems to prevent ordering.

The degree to which our particles behave as hard disks can be
estimated using the dimensionless stiffness parameter $\gamma = m_s
g/k \sigma_s$, where $g$ is the gravitational acceleration, $m_s$ is
the mass of a small particle, and $k$ is the effective spring constant
of the elastic interparticle interaction.  By measuring the
deformation of single plastic (steel) disks under gravity, we estimate
$\gamma_{\rm plastic} \approx 1.85 \times 10^{-3}$ ($\gamma_{\rm
steel} \approx 3 \times 10^{-7}$), which implies that the deviation of
the particle packings from hard-disk behavior is small (cf., Figs.\
\ref{experiment and state resolution}\subfig{b} and 
\ref{experiment and state resolution}\subfig{d}).

To generate an ensemble of frictionless MS packings, we repeatedly
performed the following protocol: The plunger was first oscillated at
high amplitude and low frequency ($50$ Hz) for $100$ ms to randomize
particle positions.  The system was allowed to relax under gravity
with the shaker turned off for $400$ ms.  We then applied a
low-amplitude, high-frequency ($400$ Hz) oscillation for $500$ ms,
which excites particle rotation and relaxes frictional
particle-particle and particle-wall interactions. Finally, the
oscillations were turned off and positions of particle centers were
determined to an accuracy of $\Delta/\sigma_s = 6\times 10^{-6}$ using
a digital camera and particle-tracking software.  The MS packings
in experiments were identified using the set of particle
positions ${\vec R}_i = \{ {\vec r}_1,{\vec r}_2,\ldots,{\vec r}_N\}$
for each configuration $i$, where ${\vec r}_j$ are the $x$- and
$y$-coordinates of the $N$ particles. (See Fig.~\ref{experiment and
state resolution}\subfig{a}.)

\paragraph{Computer Simulations}

We also performed molecular dynamics simulations of gravitational
deposition of bidisperse frictionless disks.  Our goal was to
determine how key features of MS packing distributions depend on the
particle-deposition process, and to identify which features are
robust, i.e., do not depend on specific details of the dynamics.
Thus, in the simulations we do not exactly mimic the
packing-preparation process in experiments.  In particular, we do
not model frictional contact forces, but instead we use
velocity-dependent resistance forces to dissipate energy.  However,
geometrically similar sets of MS packings are needed for a detailed
comparison of the probability distributions.  Thus, we closely match
the cell width $\lambda$, particle size distribution, gravitational
force, and elastic interactions in simulations and experiments to
obtain very similar sets of MS packings.

We assume that the disks interact via a finite-range,
purely repulsive linear spring force
\begin{equation}
\label{spring_potential}
{\vec F}^r(r_{ij})
=\frac{\epsilon}{\sigma_{ij}} \delta_{ij}
\Theta\left(-\delta_{ij}\right) {\hat r}_{ij},
\end{equation}
which mimics elastic interparticle repulsion. $\epsilon$ is
the characteristic energy scale, $r_{ij}$ is the separation between
particles $i$ and $j$, $\sigma_{ij}=\left(\sigma_i + \sigma_j
\right)/2$ is the average diameter, $\delta_{ij} = r_{ij} -
\sigma_{ij}$ is the interparticle overlap, ${\hat{r}}_{ij}$ is the
unit vector connecting particle centers, and $\Theta(x)$ is the
Heaviside step function.

To create each MS packing, we randomly place particles in a square
cell of size $L$, with no particle overlaps.  We then allow 
initially stationary particles, interacting via elastic repulsive
forces \eqref{spring_potential} and dissipative forces proportional to
relative particle velocities, to fall under gravity.  The
system evolves according to Newton's equations of motion
\begin{equation} \label{newton1}
m_i {\vec{a}}_i = -m_i g {\hat y} + 
\sum_{j\not=i}^N \left[ {\vec F}^r(r_{ij})
    - b\theta(-\delta_{ij})\vec{v}_{ij} \cdot
{\hat{r}}_{ij} \right] {\hat{r}}_{ij} + F^{w}_i,
\end{equation}
where $\vec{a}_i$ is the acceleration of particle $i$, $\vec{v}_{ij}$
is the relative velocity of particles $i$ and $j$, and $b$ is the
damping coefficient.  The particle-wall interaction force $F^{w}$ has
an analogous form to the particle-particle interaction
\eqref{newton1}, with energy scale $\epsilon^w=2 \epsilon$.  We set
the dimensionless damping coefficient to $\bar b=\sigma_s
b/\sqrt{m_s\epsilon} = 0.25$.  The simulations are terminated when the
total force ${\vec F}_{\rm tot}$ on each particle is vanishingly
small.  (In most simulations we used the threshold $F_{\rm tot} <
F_{\rm max} =10^{-14}$).

To determine which of the relaxed configurations are mechanically
stable, we calculated the eigenvalues of the dynamical matrix
\cite{tanguy}.  The MS packings possess $2N'$ positive eigenvalues, where
$N'=N-N_r$, and $N_r$ is the number of `rattler' particles.  Rattlers
have fewer than three contacts (including wall contacts)
and if present typically rest on the bottom.  In the
simulations, we distinguish distinct MS packings by comparing the
eigenvalue lists.  The eigenvalues are considered to be equal if they
differ by less than the noise threshold $10^{-6}$.  Less
than $1\%$ of the distinct MS packings contain rattlers.  In these
configurations, we ignore the translational degeneracy of the
rattlers---two configurations with the same contact networks of
non-rattler particles are treated as the same.  To compare simulation
and experimental data in configuration space, we omit the rattler
particles, and consider only positions of particles forming the
contact network.  To enumerate all MS packings and accurately measure
their frequencies, we considered small systems in the range $N=2$
to $7$ particles.  Systems with an even number of particles
contained equal numbers of large and small particles, while systems
with an odd number contained one additional small particle.

%%%%%%%%%%%%%%%%%
\begin{figure}

\pic{290}{

\putpic{-120,50}{

\put(40,0){\includegraphics[width=0.30\textwidth]
                {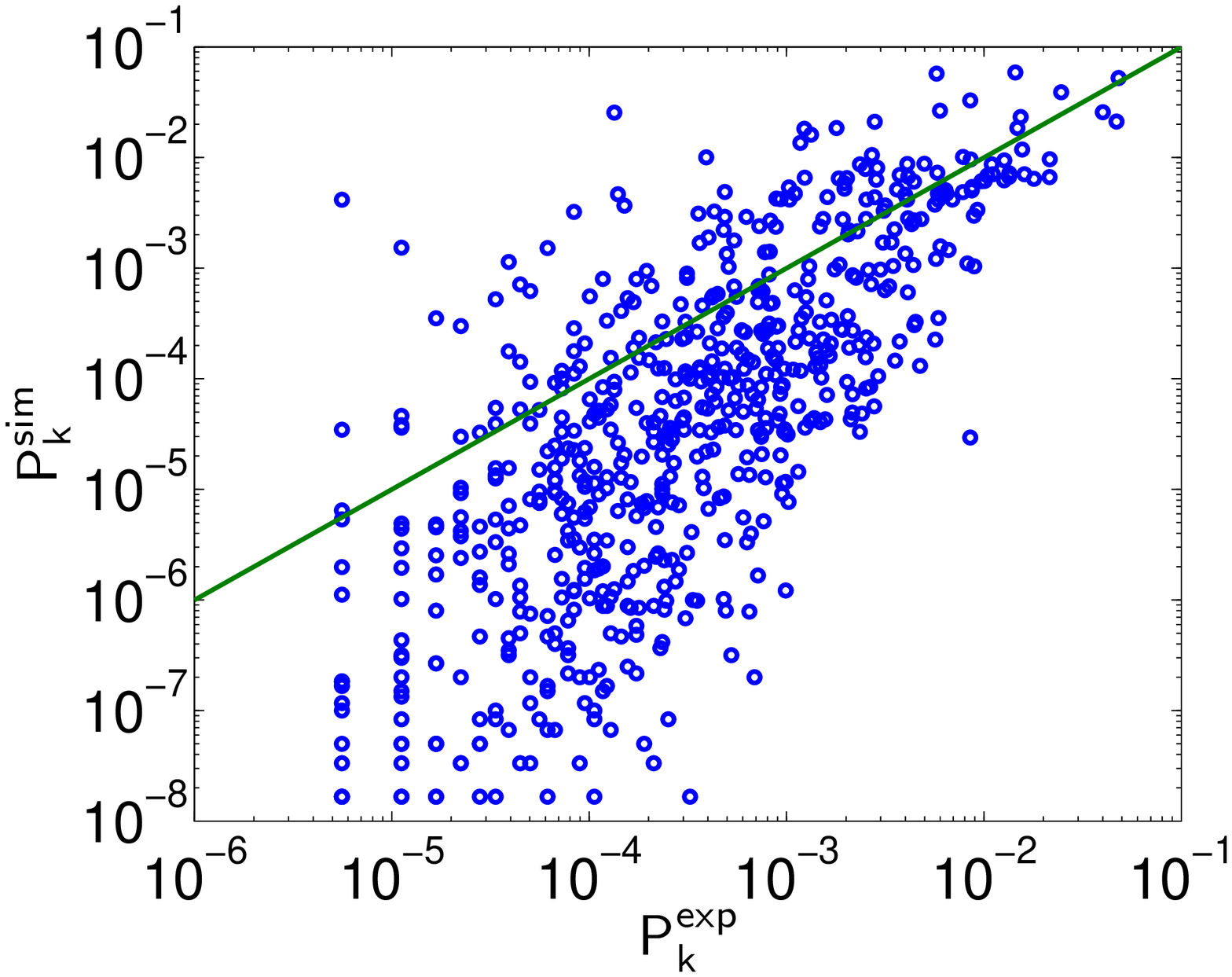}}

\put(40,120){\includegraphics[width=0.30\textwidth]
                {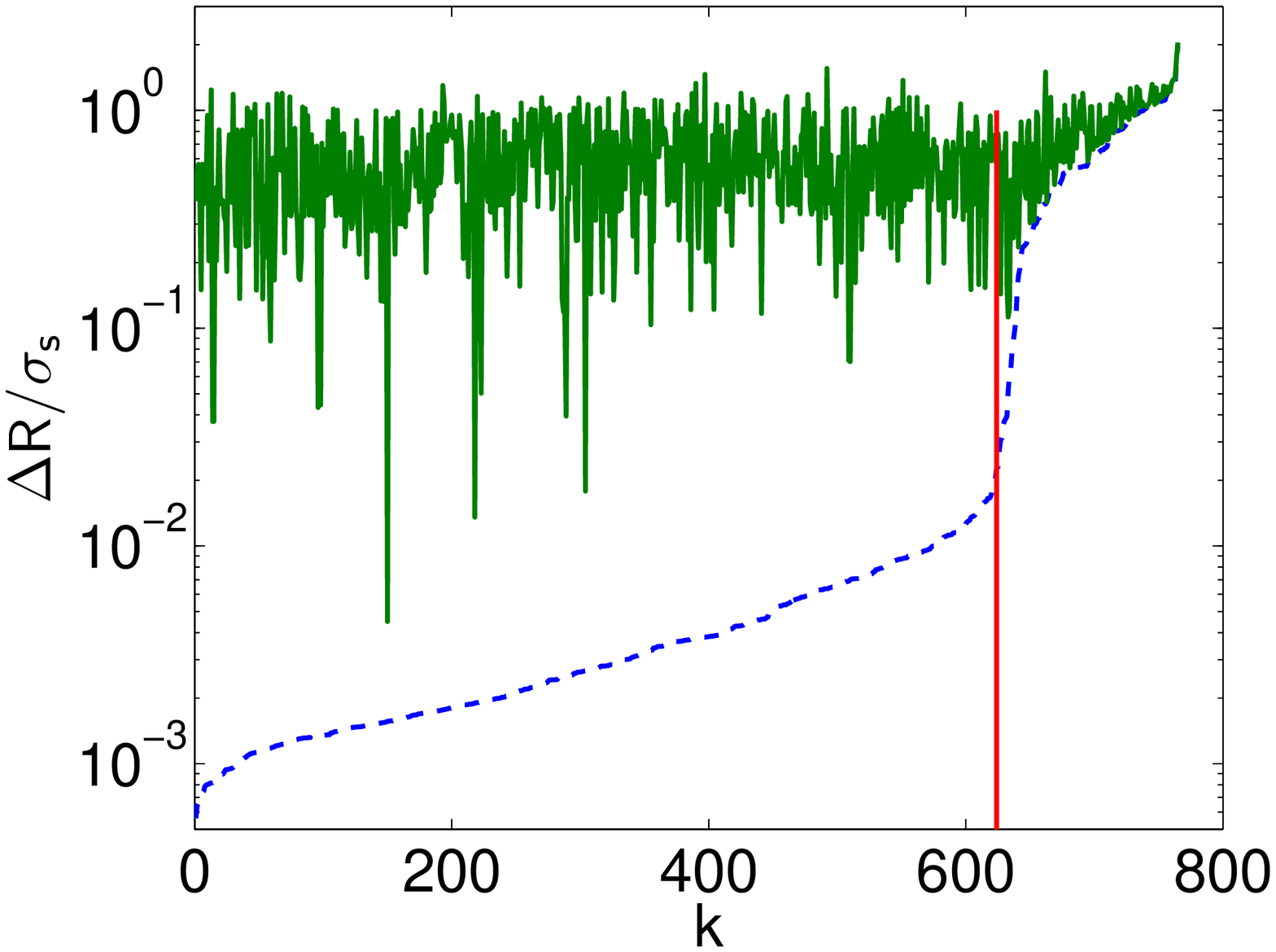}}

}
}
\vspace{-0.8in}
\caption{\label{exp-simul state separation} \subfig{a}
Distance $\Delta R_n$ and $\Delta R_{nn}$ between each MS packing
from experiments on plastic disks and the nearest
(dotted) and next-nearest (solid) MS packing
from simulations at $\gamma_{\rm plastic}$ vs.
index $k$ for the experimental MS packings sorted in order of
increasing $\Delta R_n$.  The vertical line at $k=618$ separates 
matched MS packings to the left from those that are unmatched. \subfig{b} The
probability with which MS packings occurred in experiments,
$P_k^{\rm exp}$ versus probability of the matching state in simulations
$P_k^{\rm sim}$.  The solid line has slope one.} 
\vspace{-0.2in}
\end{figure}
%%%%%%%%%%%%%%%%%  

\paragraph{Results}
We find that MS packings in frictionless granular systems occur as
discrete points in configuration space \cite{ning}.  In
Fig.~\ref{experiment and state resolution}\subfig{b} we display the
coordinates $(x_c,y_c)$ of the centroids of MS packings in a small
region containing several microstates for $N=7$ (plastic discs in
experiments and particles with $\gamma_{\rm plastic}$ in simulations).
The results show that the MS packing centroids are indeed
distinct and well-separated.  Moreover, the experimental and
simulation points agree.  Our simulations do not involve static
frictional forces, thus this agreement indicates that our novel
experimental technique is able to generate mechanically stable
frictionless packings.

Fig~\ref{experiment and state resolution}\subfig{c} shows that the
experimental and numerical scatter in the MS packing centroids is
several orders of magnitude smaller than the average separation
between discrete MS packings in configuration space.  We find that the
average distance between distinct MS packings in configuration space
is approximately $10^{-1} \sigma_s$, whereas the maximum size of the
scatter is $10^{-12} \sigma_s$ in simulations and $10^{-5} \sigma_s$
in experiments.  Based on this observation, in our analysis of
experimental data, two packings ${\vec R}_i$ and ${\vec R}_j$ are
considered to be the same microstate if $\Delta R = |{\vec R}_i - {\vec
R}_j|/\sigma_s < \Delta R^* = 0.01$.  

To determine if particles in our system can be treated as hard disks,
we tested the sensitivity of the numerically generated MS packings to
variation in the stiffness parameter $\gamma$.  The solid line in
Fig.\ \ref{experiment and state resolution}\subfig{b} shows the change
of position of the centroid of a MS packing when $\gamma$ is increased
from $10^{-7}$ to the critical value $\gamma^*$ where a sudden change
in the particle contacts occurs ($\gamma^*=0.06$ for this particular
packing).  While the overall variation of the position of the centroid
is significant, the position of the centroid for $\gamma=\gamma_{\rm
plastic}$ is essentially indistinguishable from the position in the
hard-disk limit $\gamma \rightarrow 0$.  The distribution of the
critical parameters $\gamma^*$ for the set of all simulation MS
packings is depicted in \ref{experiment and state
resolution}\subfig{d}.  These results indicate that most of the hard
sphere packings remain stable even when the stiffness parameter is
increased above $\gamma_{\rm plastic}$.

In simulations, we are able to perform an extremely large number of
trials and find nearly all MS packings in small systems.  In
Fig.~\ref{packing probabilities}\subfig{a}, we show the number of
distinct states $N_s$ as a function of the number of trials
$N_t$ for systems with $N=3$ through $7$ particles.  In all cases
(except $N=7$), we saturate the packing-generation process in the
sense that we do not generate new MS packings when the number of
trials is increased by a factor of $10$ beyond $N_t^{\rm tot}$.  In
experiments we followed a similar procedure, but we were unable to
fully saturate the curves because of insufficient number of trials.
Fig.~\ref{packing probabilities}\subfig{b} shows that in experiments
and simulations the total number of MS packings grows exponentially
with the system size, $N_s^{\rm tot} \sim \mathrm{e}^{aN}$.  The
exponent $a \approx 1.2$ is the same as found for periodic systems
\cite{ning}; however, the prefactor is larger by roughly an order of
magnitude.  The number of trials required to reach saturation of the
simulation packing-generation algorithm (cf., Fig.\ \ref{packing
probabilities}\subfig{a}) also grows exponentially with $N$, but with
a larger exponent.  Since $N_t^{\rm tot}$ grows rapidly with system
size, enumeration of $N_s^{\rm tot} \approx 728$ MS packings for $N=7$
requires $N_t^{\rm tot} \sim 10^9$ trials.  This large number
of trials stems from the extremely nonuniform packing probability 
distribution. 

The frequency distributions for MS packings are extremely nonuniform
in both simulations and experiments.  As depicted in Fig.\
\ref{packing probabilities}\subfig{c} the MS packing probabilities vary
by many orders of magnitude.  In addition, we find quantitative
agreement in the shape of the frequency distributions, which implies
that frequencies of the MS packings are only weakly sensitive to the
dynamics used to generate them.

To make a quantitative comparison between MS packings found in
experiments and simulations, we calculated the distance
in configuration space $\Delta R$ between each MS packing generated in
experiments and the nearest and next nearest MS packings found in
simulations.  In Fig.~\ref{exp-simul state separation}\subfig{a}, we
show the nearest-neighbor and next-nearest neighbor separations
$\Delta R_n$ and $\Delta R_{nn}$ for experiments with $N=7$ plastic
disks and the corresponding simulations at $\gamma_{\rm plastic}$
versus index $k$ sorted by increasing $\Delta R_n$.  For
approximately $80$\,\% of the packings the nearest-neighbor and
next-nearest-neighbor distances are well separated, with $\Delta R_n$
much smaller than the average distance between packings shown in Fig.\
\ref{experiment and state resolution}\subfig{c}.  This separation of
length scales in configuration space allows an unambiguous match
between packings found in experiments and simulations.  At $k\approx
k^*=618$ the distance $\Delta R_n$ rapidly increases, and packings
with $k > k^*$ are unmatched.

We find that $n=110$ ($m=146$) MS packings from simulations at
$\gamma_{\rm plastic}$ (experiments on plastic disks) are unmatched
for $N=7$.  However, approximately $70$\,\% of the unmatched experimental
packings are unstable when used as initial conditions in the
simulations. (In experiments, these packings are likely stabilized
by residual frictional forces or small inaccuracies in the numerical
representation of the experiment.)  For example, the results
shown in Fig.\ \ref{experiment and state resolution}\subfig{d}
indicate that stability of some packings may depend on the stiffness
parameter near $\gamma\approx\gamma_{\rm plastic}$.  The unmatched MS
packings from simulation most likely result from insufficient
experimental statistics.

To determine the sensitivity of packing probability distributions
on the particle deposition process, in Fig.~\ref{exp-simul state
separation}\subfig{b} we compare the experimental probabilities
$P_k^{\rm exp}$ with the probabilities found in simulations $P_k^{\rm
sim}$ for the set of matched states.  We demonstrate a strong
correlation between $P_k^{\rm exp}$ and $P_k^{\rm sim}$: likely
packings in experiments tend to be likely in simulations, and rare
packings in experiments tend to be rare in simulations (although there
is also a significant scatter).  In fact, we calculate that the rms
deviation in the probabilities of matched packings in simulations and
experiments is only $8\%$ of the probability of the most frequent MS
packing.  Since the dynamics in experiments and simulations is quite
different, this is an important result, which implies that properties
of static frictionless packings are weakly dependent on the
packing-preparation protocol.

\paragraph{Conclusions}
We introduced a novel experimental method to generate frictionless MS
packings of granular materials.  This method is crucial for studies
aimed at differentiating the effects of geometrical constraints and
friction on the structural and mechanical properties of jammed
granular systems.  We performed coordinated experimental and
computational studies of frictionless MS packings in small systems,
which showed that MS packing probabilities are extremely nonuniform
and relatively insensitive to the procedure used to prepare them.  In
future studies, we will investigate the consequences of our present
results for the microstate statistics in macroscopic granular systems
treated as a collection of nearly independent small subsystems.  We
will also dial in frictional contacts to generate continuous
geometrical families of MS packings that occur even at fixed $\gamma$,
and then compare the statistics of these continuous sets of packings
to that for discrete MS packings.
    
\begin{acknowledgments}
Financial support from NSF grant nos. CBET-0348175 (GG, JB) and
DMR-0448838 (GG, CSO) is gratefully acknowledged.  MS thanks Yale
University for lab and office space during his sabbatical when this
work was performed.  

\end{acknowledgments}

\end{document}